\documentclass{article}

\usepackage[margin=1in]{geometry} 
\usepackage{amsmath,amsthm,amssymb}
\usepackage[english]{babel}
\usepackage{caption}
\usepackage{subcaption}
\usepackage{tikz-cd}
\usepackage{enumitem}
\setlist[itemize]{nolistsep}
\setlist[enumerate]{label=(\alph*),nolistsep}
\usepackage{hyperref}
\usepackage{bbm}
\usepackage{bm}
\usepackage{physics}
\usepackage{outlines}
\usepackage[textsize=small]{todonotes}
\usepackage{multirow}

\DeclareMathOperator*{\argmax}{arg\,max}

\begin{document}
\title{Efficient magnetometer sensor array selection for signal reconstruction and brain source localization}
\author{\normalsize{Wan-Jin Yeo$^{1,2}$, Samu Taulu$^{1,2}$, and J. Nathan Kutz$^{1,3}$}\\
\footnotesize{$^1$ Department of Physics, University of Washington, Seattle, WA 98195}\\
\footnotesize{$^2$ Institute for Learning and Brain Sciences, University of Washington, Seattle, WA 98195}\\
\footnotesize{$^3$ Department of Applied Mathematics, University of Washington, Seattle, WA 98195}\vspace{-.0in}}

\date{\today}
\maketitle

\begin{abstract}
Magnetoencephalography (MEG) is a noninvasive method for measuring magnetic flux signals caused by brain activity using sensor arrays located on or above the scalp.
A common strategy for monitoring brain activity is to place sensors on a nearly uniform grid, or sensor array, around the head.
By increasing the total number of sensors, higher spatial-frequency components of brain activity can be resolved as dictated by Nyquist sampling theory.
Currently, there are few principled mathematical architectures for sensor placement aside from Nyquist considerations.
However, global brain activity often exhibits low-dimensional patterns of spatio-temporal dynamics.
The low-dimensional global patterns can be computed from the singular value decomposition and can be leveraged to select a small number of sensors optimized for reconstructing brain signals and localizing brain sources.
Moreover, a smaller number of sensors which are systematically chosen can outperform the entire sensor array when considering noisy measurements. 
We propose a greedy selection algorithm based upon the QR decomposition that is computationally efficient to implement for MEG. %
We demonstrate the performance of the sensor selection algorithm for the tasks of signal reconstruction and localization. 
The performance is dependent upon source localization, with shallow sources easily identified and reconstructed, and deep sources more difficult to locate.
Our findings suggest that principled methods for sensor selection can improve MEG capabilities and potentially add cost savings for monitoring brain-wide activity.
\end{abstract}

\section{Introduction} \label{introduction}

Magnetoencephalography (MEG) is a noninvasive method that uses sensors located on or above the scalp to record magnetic flux signals caused by brain activity \cite{hamalainen93,hari17,ilmoniemi19}. There are mainly two MEG sensor types currently in use: the first is the more common {\em superconducting quantum interference device} (SQUID) sensor, which measures the surface magnetic flux through gradiometer, or magnetometer pick-up loops. SQUID sensors require cryogenic temperatures to work, and hence are placed in a dewar filled with liquid helium that surrounds the head. These dewars are necessarily thick-walled to ensure thermal insulation, and so the closest distance between the sensors and the scalp is limited to around 2 cm. The second type of MEG sensor is the optically-pumped magnetometer (OPM), which measures volumetric flux through a cylindrical sensing volume along a sensing direction that can be modulated. They are the proposed next-generation sensor, and may be placed directly onto the scalp since they can be operated at room temperatures. In practice, both SQUID and OPM are used in sensor arrays placed around the head. To date, there have been some discussions into using a limited number of sensors \cite{nenonen22,riaz17}, but only few methods exist to evaluate which sensors are most capable of reconstructing brain-wide activity or source localization \cite{beltrachini21, iivanainen21}. We demonstrate that by exploiting low-dimensional patterns of brain activity along with a greedy selection algorithm, we can identify a small number of sensors that can accurately reconstruct brain signals and source localization.

MEG sensor array design has predominantly been driven by the application of Nyquist sampling theory, which dictates that higher spatial frequencies can be resolved by decreasing the distance between sensors. There is thus a general consensus that an increase in the total number of sensors can improve signal resolution; a higher array density allows us to detect these higher spatial frequency components and hence prevent signal aliasing \cite{ahonen93,boto19,tierney20}. This appears especially relevant for next-generation OPM sensor arrays, which can be placed directly on the head. In theory, since higher spatial frequency components decay quicker than lower spatial frequency components, decreasing sensor-to-source distance means the higher frequency components can now be detected due to the higher signal-to-noise ratio (SNR) \cite{taulu05,boto16,iivanainen17,iivanainen19}. Other motivations for an increased number of sensors include advantages in source localization accuracy in the beamforming framework \cite{vrba04} and the ability to extract more information from the signal in general which is motivated by Shannon information theory \cite{kemppainen89,nenonen07}. However, having denser sensor arrays presents a few problems; first, the available sensor region may be unable to accommodate sufficient sensors physically. Secondly, it may be costly to manufacture such a large amount of sensors. The ability to select just a few optimal sampling locations via exploiting the low-dimensional representation of signals will alleviate these potential problems. Moreover, it allows us to perform non-compromised testing during developmental phases where only limited sensors are available.

%  As sensors are placed closer to the head, a higher sensor density may be required, since the closer sensor array is expected to be able to measure higher spatial frequency signal components.   As a consequence, SQUID and OPM systems require different sensor array strategies because of the placement on (OPM) or near (SQUID) the head.

Apart from sensor array density, MEG sensor design can also be improved with informed sensor selection techniques. Current sensor layouts for MEG are primarily guided by physical/hardware considerations \cite{hill20}, or simply made to sample the head uniformly in some way \cite{nenonen22,riaz17}. In fact, a hexagonal sampling lattice has be shown to be the optimally sparse sampling grid under the above mentioned Nyquist sampling consideration \cite{ahonen93}. Sensor selection is able to determine where to optimally place sensors in an informed and principled manner based on the neural activity itself. This allows for an alternate optimal sensor design that is tailored based on actual data, which may be more effective than utilizing a universal, uniformly-spaced sensor design. Recently, there have been discussions that suggest that since some OPM sensor arrays allow for flexible sensor position placement, it may allow for fewer but more effectively-placed sensors \cite{marhl22,iivanainen21}. Our results indicate this to be true.

\begin{figure}[t]
    \centering
        \includegraphics[width=\textwidth]{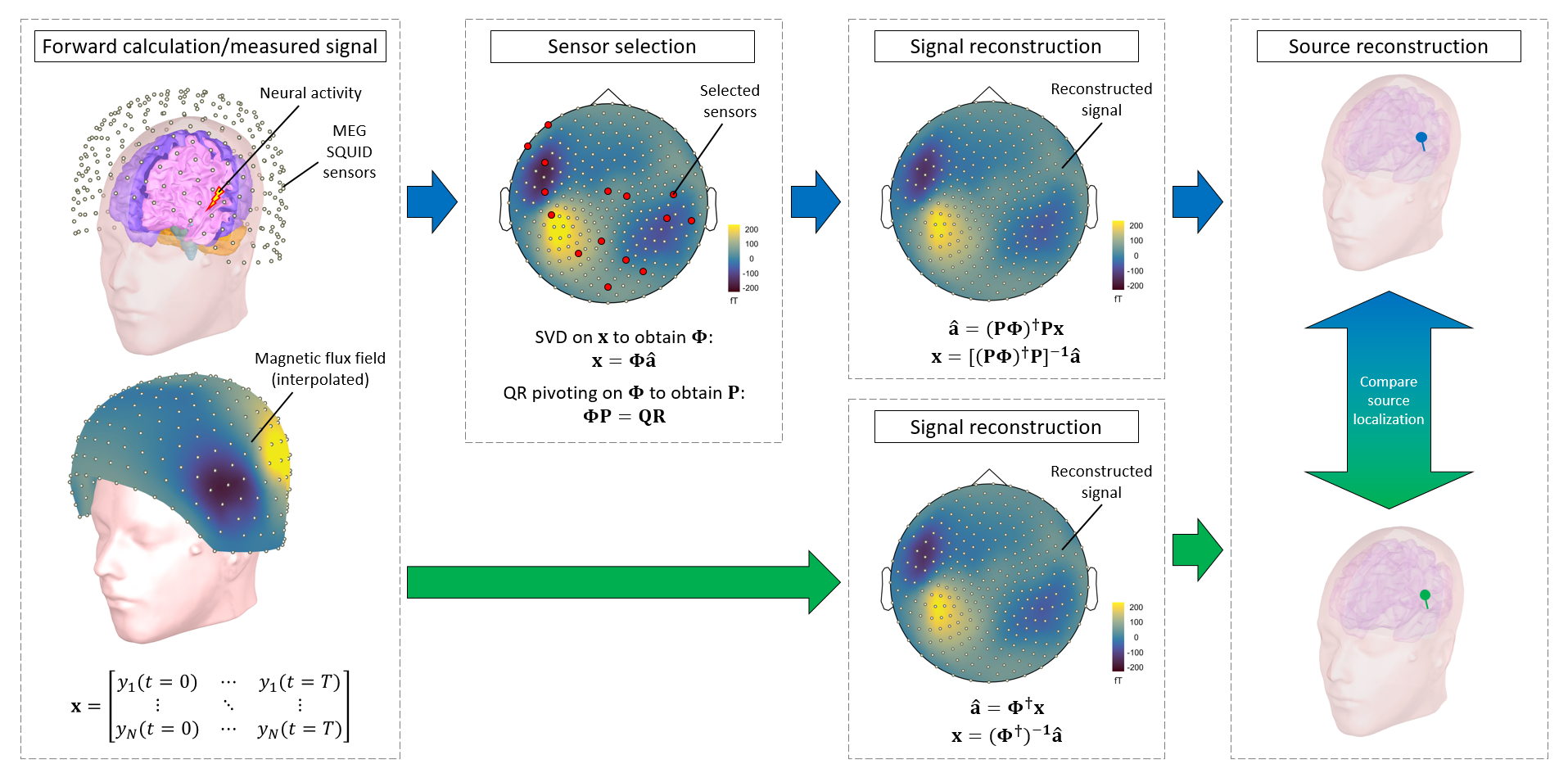}
        \caption{The green arrows indicate the usual pipeline for signal and source reconstruction that utilizes all sensors on the sensor array. For sparse sampling, we propose using a principled sensor selection using QR pivoting on a sparse basis representation obtained via SVD, as shown in the sequence of blue arrows. We then only require a subset of sensors for signal reconstruction and source localization. Here, $\mathbf{x}$ denotes the full-state signal measurement, $\bm{\Phi}$ denotes the rank-$r$ truncated left singular basis obtained via SVD of $\mathbf{x}$, $\hat{\mathbf{a}}$ denotes the (noisy) coefficients for modes of $\bm{\Phi}$, and $\mathbf{P}$ denotes the permutation matrix to select rows/sensors of $\bm{\Phi}$. These notations are defined throughout Section 2.}
        \label{fig:1}
\end{figure}

In this paper, we apply the QR pivoting algorithm~\cite{trefethen1997numerical,brunton19,manohar18} on the left singular $r$-rank truncated SVD basis of MEG SQUID data to determine $p=r$ optimal sensor locations. QR pivoting is a greedy selection algorithm that maximizes a matrix's determinant by identifying and selecting sensor locations exhibiting maximal variance. The order of sensors selected determine their importance in representing the basis modes and accurate signal reconstruction. This procedure, which also permutes the rows of the basis matrix based on the sensor selection order, will equivalently result in the minimization of an error covariance matrix measuring signal reconstruction error as desired. We use the SVD basis since it reveals a latent, low-dimensional representation of the signal, indicating the redundancy of having a larger number of sensors beyond that required to represent the small number of significant modes. A summary of the procedure utilizing QR pivoting is shown in Figure~\ref{fig:1} alongside the traditional approach of signal reconstruction and source localization.

QR has been widely used in applications including fluid flows, sea surface temperature monitoring, and face image recognition \cite{manohar18,clark18,jayaraman20,gudivada12}. It is computationally efficient and is one of the leading paradigms for monitoring engineering and physical systems. The sensor selection procedure is demonstrated on two example datasets that are available on the Brainstorm website (\url{http://neuroimage.usc.edu/brainstorm}).

\section{Theoretical background}

The theoretical underpinnings of the QR method and its relation to sensor selection are summarized in what follows. Specifically, we provide a mathematical construction of the problem statement, which is followed by a review of the {\em singular value decomposition} (SVD), the signal reconstruction, and the QR selection algorithms.

\subsection{Problem statement}
Our goal is to select a subset of optimal sensors that is able to sparsely sample the signal, yet still be able to reconstruct the full signal adequately well. Let the full signal be $\mathbf{x} \in \mathbb{R}^{n\times m}$, where $n$ is the total number of sensors, and $m$ is the number of temporal samples. Also let $\bm{\eta} \in \mathbb{R}^{n \times m}$ be an additive independent identically distributed Gaussian noise $\mathcal{N} (0 , \sigma^2)$. For MEG, entries of $\mathbf{x}$ correspond to magnetic flux measured through each sensor. The signal may be represented by some basis $\bm{\Phi} \in \mathbb{R}^{n \times k}$,
\begin{align} \label{eq: basic_eqn}
    \mathbf{x} = \bm{\Phi} \mathbf{a} + \bm{\eta} \equiv \bm{\Phi} \hat{\mathbf{a}}
\end{align}
where $k$ is the number of basis vectors/modes, and each column of $\mathbf{a} \in \mathbb{R}^{k \times m}$ contain the coefficients to each basis mode. We have also defined 
\begin{align}
    \hat{\mathbf{a}} \equiv \mathbf{a} + \bm{\Phi}^{\dagger} \bm{\eta}
\end{align}
as the ``noisy'' coefficient/weights, where $^\dagger$ denotes the Moore–Penrose pseudoinverse~\cite{barata2012moore,golub2013matrix} (in this case, the right pseudoinverse). The goal is to find a matrix $\mathbf{P} \in \mathbb{R}^{p \times n}$ that picks out $p < n$ number of sensors to sample the signal so that we may still reconstruct the full signal $\mathbf{x}$ accurately. In other words, we want to be able to obtain a full-state reconstruction from a reduced-state signal
\begin{align}\label{eq: basic_eqn_red}
    \mathbf{y} = \mathbf{P}\bm{\Phi} \hat{\mathbf{a}} \equiv \bm{\Theta} \hat{\mathbf{a}} = \bm{\Theta} \mathbf{a} + \mathbf{P} \bm{\eta}
\end{align}
where we have defined $\bm{\Theta} = \mathbf{P} \bm{\Phi}$. Evidently, $\mathbf{P}$ must be a permutation matrix with rows of canonical unit vectors that select the appropriate rows of $\bm{\Phi}$. 

The QR pivoting algorithm, which is a greedy determinant-maximizing algorithm, approximately provides a solution to this problem and will be discussed in Section~\ref{qr_pivoting_algorithm}.
Alternative sparse selection algorithms can be constructed, including a relaxed formulation which can potentially provide even better performance~\cite{zheng2018unified}.  And for very high-dimensional problems, randomized algorithms~\cite{erichson2016randomized} and compressed decompositions~\cite{erichson2019compressed} can be exploited.  However, the QR algorithm provides exceptional performance with minimal computational cost.

\subsection{Low-rank subspaces and SVD}

Many datasets and signals exhibit low-dimensional patterns in certain basis choices that may be exploited for easier computation. One such basis may be found via the SVD. SVD decomposes some matrix $\mathbf{A} \in \mathbb{R}^{a \times b}$ into unitary matrices $\mathbf{U} \in \mathbb{R}^{a \times a}$ and $\mathbf{V} \in \mathbb{R}^{b \times b}$, and a diagonal matrix $\mathbf{S} \in \mathbb{R}^{a \times b}$ with non-negative entries (singular values),
\begin{align}
    \mathbf{A} = \mathbf{USV}^\dagger
\end{align}
where the singular values of $\mathbf{S}$ are arranged in decreasing order.

By the Eckart-Young theorem \cite{eckart36,brunton19}, if we approximate $\mathbf{A}$ by truncating $\mathbf{U}$ and $\mathbf{V}$ of its SVD to the first $r$ columns, then it is the optimal rank-$r$ least-squares approximation of $\mathbf{A}$. In other words, the columns/modes of $\mathbf{U}$ and $\mathbf{V}$ with higher corresponding singular values have higher dominance in representing the entire signal. If a signal's SVD has only few significant singular values, then it can be well-approximated by just the first few modes of $\mathbf{U}$ and $\mathbf{V}$, reducing the signal from a high-dimensional space to a low-dimensional subspace.

Note that when $a>b$ (overdetermined), the lower $(a-b)\times b$ submatrix of $\mathbf{S}$ is necessarily null. Hence, we only need to consider the first $b$ columns of $\mathbf{U}$ and $\mathbf{V}$. When $a<b$ (underdetermined), the rightmost $a \times (b-a)$ submatrix of $\mathbf{S}$ is null, hence we only need to consider the first $a$ rows of $\mathbf{V}$. In other words, only square submatrix of $\mathbf{S}$ with non-negative diagonal is kept. These two cases are called ``economy SVD'' and are utilized to simplify computation~\cite{kutz2013data}.

We will use economy SVD for the rest of the paper, with only the first 
$r \leq b$ or $r\leq a$ significant modes preserved for the overdetermined and underdetermined cases respectively. In the context of equation \eqref{eq: basic_eqn} we have $\mathbf{A} = \mathbf{y}$, $a=n$, $b=m$, so using the basis choice of the rank-$r$ truncated $\mathbf{U}$ means that we have $\mathbf{U} = \mathbf{\Phi} \in \mathbb{R}^{n \times r}$ (with number of rows greater or equal to number of columns) and $\hat{\mathbf{a}} = \mathbf{SV}^{\dagger} \in \mathbb{R}^{r \times m}$.

\subsection{Sensor selection and signal reconstruction}

\subsubsection{Coefficient estimation and signal reconstruction}
We now assume that we have selected a total of $p \geq r$ sensors; the QR algorithm to do so is explained in Section~\ref{qr_pivoting_algorithm}. That is, we have identified $r$ $n$-dimensional patterns that are used to project our $p$ sensor measurements. Each pattern corresponds to, for instance, the spatial topography of a specific brain response. If such a pattern were generated by a single current dipole in the cortex, the corresponding number of parameters to be determined from the pattern would be 6 (dipole location and orientation). The optimal number of sensors for the best sparse measurement is generally not trivial; however, discussions in \cite{clark20,manohar18} suggest that the optimal number may be approximately the underlying rank of the system. This is supported by the relative error plots in Figures~\ref{fig:phantom_relerr} and \ref{fig:binaural_err} of our paper as well. As such, we do not consider the $p<r$ case here; we want at least $p=r$ chosen sensors so that all $r$ basis modes are best represented by the $p\geq r$ sensors and hence useful. A discussion of the underdetermined case where $p<r$ can be found in \cite{saito21}.

In order to reconstruct the noiseless signal, we need to obtain the unknown coefficients $\mathbf{a}$. However, only the noisy coefficient $\hat{\mathbf{a}}$ can be obtained as an estimate of $\mathbf{a}$; its least squares solution can be found by taking the (pseudo)inverse of $\bm{\Theta}$,
\begin{align}
    \bm{\Theta}^\dagger \mathbf{y} = \hat{\mathbf{a}} = \mathbf{a} + \mathbf{\Phi}^{\dagger} \bm{\eta} =
    \begin{cases}
    \bm{\Theta}^{-1} \mathbf{y}, \quad p = r \\
    \left(\bm{\Theta}^T \bm{\Theta}\right)^{-1} \bm{\Theta}^T \mathbf{y}, \quad p > r.
    \end{cases}
\end{align}
With the estimate $\hat{\mathbf{a}}$, the reconstructed full-state signal $\mathbf{x}$ can then be obtained from \eqref{eq: basic_eqn}.

To quantify the estimation error, we consider the error covariance matrix Var$[\mathbf{a} - \hat{\mathbf{a}}]$. For accurate noiseless signal reconstruction, the volume, or equivalently, the absolute determinant, of this error covariance matrix should be minimized. A direct calculation shows that
% \begin{align} \label{variance}
%     \text{Var}\left[\mathbf{x} - \mathbf{z}\right] = \begin{cases}
%     \sigma^2 \left(\bm{\Phi} \bm{\Phi}^T \right)^{-1}, \quad p < m \\
%     \sigma^2 \left(\bm{\Phi}^T \bm{\Phi} \right)^{-1}, \quad p \geq m
%     \end{cases} 
% \end{align}
\begin{align} \label{eq: variance}
    \text{Var}\left[\mathbf{a} - \hat{\mathbf{a}}\right] = \sigma^2 \left(\bm{\Phi}^T \bm{\Phi} \right)^{\dagger} = \sigma^2 \left(\bm{\Theta}^T \bm{\Theta} \right)^{-1} , \quad p \geq r
\end{align}
where we have used the fact that $\mathbf{\Phi}^\dagger = \bm{\Theta}^\dagger \mathbf{P}$ from the definition of $\bm{\Theta}$. In other words, to minimize the absolute determinant of the covariance matrix, we may equivalently maximize the determinant of
% $\bm{\Phi}^T \bm{\Phi}$ or
$\bm{\Theta}^T \bm{\Theta}$ (due to the inverse). This may be done by permuting its rows in an optimal manner using the pivoted QR algorithm as discussed in Section~\ref{qr_pivoting_algorithm}; this corresponds to ranking sensors that best represent the basis modes. Compactly, we may write this as finding an index set $\gamma = \{ \gamma_1, \dots, \gamma_p \} \subseteq \{ 1, \dots, n\}$ such that
\begin{align} \label{eq: goal_pgeqm1}
    \gamma^* = 
    % \argmax_{\gamma, \abs{\gamma} = p} \abs{ \det \left( \bm{\Phi}^T_\gamma \bm{\Phi}_\gamma \right)} = 
    \argmax_{\gamma, \abs{\gamma} = p} \abs{ \det \left( \bm{\Theta}^T_\gamma \bm{\Theta}_\gamma \right)}, \quad p \geq r
\end{align}
where 
% $\bm{\Phi}_\gamma$ denotes the permuted basis matrix, and
$\bm{\Theta}_\gamma = \mathbf{P}_\gamma \bm{\Phi}$ is the appropriate permutation matrix. Since $\bm{\Theta}^T \bm{\Theta}$ is square, its determinant is the product of its eigenvalues. Moreover, $\bm{\Theta}^T \bm{\Theta}$ is symmetric positive semidefinite, so its eigenvalues coincide with its singular values $\sigma$. We know that the first $r$ eigenvalues/singular values of $\bm{\Theta}^T \bm{\Theta}$ and $\bm{\Theta} \bm{\Theta}^T$ are identical as well. Thus, \eqref{eq: goal_pgeqm1} is equivalently
\begin{align}
    \gamma^* &= \argmax_{\gamma, \abs{\gamma} = p} \abs{ \prod_{i=1}^r \sigma_i (\bm{\Theta}_\gamma^T \bm{\Theta}_\gamma)} = \argmax_{\gamma, \abs{\gamma} = p} \abs{\prod_{i=1}^r \sigma_i (\bm{\Theta}_\gamma \bm{\Theta}_\gamma^T)} \\
    &= \argmax_{\gamma, \abs{\gamma} = p} \abs{ \det \left( \bm{\Theta}_\gamma \bm{\Theta}_\gamma^T \right)}, \quad p \geq r \label{eq: goal_pgeqm}
\end{align}
Note that the parentheses on the second and third equalities do not indicate multiplication, but instead they indicate that the singular values $\sigma$ are that of $\bm{\Theta}^T \bm{\Theta}$ and $\bm{\Theta} \bm{\Theta}^T$ respectively.

% Moreover, $\mathbf{A}^T \mathbf{A}$ and $\mathbf{A} \mathbf{A}^T$ are symmetric and positive semidefinite, so their eigenvalues are the squares of the singular values of $A$.

In the $p=r$ case where $\bm{\Theta}$ is square, we know that $\det \bm{\Theta}^T = \det \bm{\Theta}$, so $\det(\bm{\Theta}^T \bm{\Theta}) = \det \bm{\Theta}^T \det \bm{\Theta} = (\det \bm{\Theta})^2$. We thus only need to maximize the determinant of $\bm{\Theta}$,
\begin{align} \label{eq: goal_peqm}
    \gamma^* = \argmax_{\gamma, \abs{\gamma} = p} \abs{ \det \bm{\Theta}_\gamma} , \quad p = r
\end{align}

\subsubsection{QR pivoting algorithm} \label{qr_pivoting_algorithm}

The pivoted QR algorithm/decomposition was first introduced in \cite{businger65}, but its application towards greedy sensor selection has only been done recently in \cite{drmac16}. It follows the usual QR algorithm that decomposes some matrix $\mathbf{A} \in \mathbb{R}^{a \times b}$ into a unitary matrix $\mathbf{Q} \in \mathbb{R}^{a \times a}$ and an upper triangular matrix $\mathbf{R} \in \mathbb{R}^{a \times b}$, except that at each iterative step, the column with the largest 2-norm is first selected and then swapped with the first column. The pivoted QR decomposition hence includes an appropriate column permutation matrix $\mathbf{P} \in \mathbb{R}^{b \times b}$,
\begin{align}
    \mathbf{A} \mathbf{P} = \mathbf{Q} \mathbf{R}
\end{align}

Consider the first iteration. The first step as mentioned is to select the column of $\mathbf{A}$ with the largest 2-norm, and swapping it with the first column. Assuming the column with largest 2-norm is the $k$\textsuperscript{th} column, this is equivalent to right multiplying $\mathbf{A}$ by a permutation matrix 
\begin{align}
    \mathbf{P}_1 = \begin{bmatrix}
        \mathbf{e}_k \ \mathbf{e}_2 \ \dots \ \mathbf{e}_{k-1} \ \mathbf{e}_1 \ \mathbf{e}_{k+1} \ \dots \ \mathbf{e}_n
    \end{bmatrix}
\end{align}
where $\mathbf{e}_{k} \in \mathbb{R}^{b \times 1}$ denotes the $k$\textsuperscript{th} canonical basis vector, and $\mathbf{P}_1$ denotes permutation matrix for the first iteration of the QR pivoting algorithm. After the swap, we apply a Householder transformation on the first column of $\mathbf{A}$. Let $\mathbf{b}_1 \equiv \mathbf{a}_1 - \norm{\mathbf{a}_1 }\mathbf{e}_1$ where $\mathbf{a}_1$ is the first column of $\mathbf{A}$, then the Householder transformation looks like
\begin{align}
    \mathbf{Q}_1 = \mathbf{I} - \frac{2 \mathbf{b}_1\mathbf{b}_1^*}{\norm{\mathbf{b}_1}^2}
\end{align}
where $\mathbf{I} \in \mathbb{R}^{a \times a}$ is the identity matrix. The Householder transform corresponds to a reflection of a vector with respect to its orthogonal subspace. A straightforward calculation shows that 
\begin{align} \label{eq: householder}
    \mathbf{Q}_1 \mathbf{A} \mathbf{P}_1 =
    \begin{bmatrix}
        \norm{\mathbf{a}_k} & * & \dots & * \\ 
        0 & & &  \\
        \vdots & & \mathbf{A}^{(2)} & \\
        0 & & & 
    \end{bmatrix}
\end{align}
where $\mathbf{A}^{(2)} \in \mathbb{R}^{(a-1)\times(b-1)}$ is the submatrix of $\mathbf{A}$ excluding the first row and column. We denote a $\mathbf{A}^{(k)}$ as the bottom right $(a-k+1)\times(b-k+1)$ submatrix of $\mathbf{A}$ (so $\mathbf{A} = \mathbf{A}^{(1)}$).

We then repeat the same process on $\mathbf{A}^{(2)}$ and subsequent submatrices until $\mathbf{A}$ is decomposed into an upper triangular matrix; applying the Householder transformation for subsequent submatrices at every iteration guarantees we will reach an upper triangular matrix form due to \eqref{eq: householder}. In the case of $a>b$, this is achieved after $b$ iterations, and in the case of $a\leq b$, it is achieved after $a$ iterations. Note that in order for the Householder transformation of the $k$\textsuperscript{th} iteration to act on the appropriate first column of $\mathbf{A}^{(k)}$, and for the permutation matrix to act on just columns of $\mathbf{A}^{(k)}$, they resemble the form 
\begin{align}
    \mathbf{Q}_{k} = \begin{bmatrix}
        \mathbf{I} & 0 & \dots & 0 \\ 
        0 & & &  \\
        \vdots & & \mathbf{Q}^{(k)}_k & \\
        0 & & & 
    \end{bmatrix}, \qquad \mathbf{P}_k = \begin{bmatrix}
        \mathbf{I} & 0 & \dots & 0 \\ 
        0 & & &  \\
        \vdots & & \mathbf{P}^{(k)}_k & \\
        0 & & & 
    \end{bmatrix}
\end{align}
where $\mathbf{I} \in \mathbf{R}^{(k-1) \times (k-1)}$. For $a\leq b$, the final result will be
\begin{align}
    \mathbf{Q}_a \dots \mathbf{Q}_2 \mathbf{Q}_1 \mathbf{A} \mathbf{P}_1 \mathbf{P}_2 \dots \mathbf{P}_a =  \begin{bmatrix}
        \norm{\mathbf{a}_{k_1}^{(1)}} & * & * & * & * & \dots & * \\
        0 & \norm{\mathbf{a}_{k_2}^{(2)}} & * & * & * & \dots & * \\
        0 & 0 & \ddots & * & * & \dots & * \\
        0 & 0 & 0 & \norm{\mathbf{a}_{k_m}^{(a)}} & * & \dots & *
    \end{bmatrix} 
\end{align}
Therefore, 
\begin{align}
    \mathbf{R} &= \mathbf{Q}_a \dots \mathbf{Q}_2 \mathbf{Q}_1 \mathbf{A} \mathbf{P}_1 \mathbf{P}_2 \dots \mathbf{P}_a \\
    \mathbf{P} &= \mathbf{P}_1 \mathbf{P}_2 \dots \mathbf{P}_a \\
    \mathbf{Q} &= \left(\mathbf{Q}_a \dots \mathbf{Q}_2 \mathbf{Q}_1 \right)^\dagger
\end{align}

From the above description, we see that the pivoted QR algorithm greedily maximizes the determinant of $\mathbf{R} (1:a, 1:a)$ (i.e. its first $a\times a$ submatrix) since a diagonal dominant structure is constructed. The diagonal entries are the greedily-obtained maximum column 2-norm of the (sub)matrices of $\mathbf{A}$, which maximizes determinant since the determinant of a triangular matrix is the product of its diagonal elements. Moreover, note that
\begin{align} \label{eq: reason_peqm}
    \abs{\det \mathbf{R}(1:a,1:a)} = \abs{\det \left(\mathbf{Q}^\dagger \mathbf{A} \mathbf{P}(:,1:a) \right)} = \abs{\det \left(\mathbf{A} \mathbf{P} (:,1:a)\right)}
\end{align}
since $\mathbf{Q}^\dagger$ is unitary so $\abs{\det \mathbf{Q}^\dagger} = 1$. For the $a=b$ case, we will have the even further simplified case
\begin{align} \label{eq: reason_pgeqm}
    \abs{\det \mathbf{R}} = \abs{\det \left(\mathbf{Q}^\dagger \mathbf{A} \mathbf{P} \right)} = \abs{\det \mathbf{A}}
\end{align}
since for (square) permutation matrices, $\abs{\det \mathbf{P}} = 1$.

In our context, the goal is is to permute the sensors/rows of either $\bm{\Theta} \bm{\Theta}^T$ or $\bm{\Theta}$ for the $p>r$ case or $p=r$ case as in \eqref{eq: goal_pgeqm} and \eqref{eq: goal_peqm} respectively. For maximum sparsity, we consider only the $p=r$ case in this paper so that a minimum number of sensors are required to sample the $r$ modes. We thus perform QR pivoting on $\mathbf{A} = \bm{\Phi}^T$ (transposed since QR pivoting acts on columns) so that eventually, $\abs{\det \bm{\Phi}^T(1:r,1:r)}$ as in \eqref{eq: reason_peqm} is greedily maximized, satisfying the requirement \eqref{eq: goal_peqm} where only the first $p=r\leq n$ is selected.

For the $p > r$ case, one way to approximately satisfy \eqref{eq: goal_pgeqm} is to maximize the determinant of $\bm{\Phi} \bm{\Phi}^T$. Performing QR pivoting on $\mathbf{A} = \bm{\Phi} \bm{\Phi}^T$ greedily maximizes \eqref{eq: reason_pgeqm}. However, it is shown in \cite{saito21} that this is in fact not optimized to solve \eqref{eq: goal_pgeqm}, but only provides a somewhat informed sensor selection that still has better performance than random sensor selection. In the same paper \cite{saito21}, an improved extended algorithm for sensor selection in the $p > r$ case is provided.

\section{Results}

To evaluate the QR pivoting sensor selection algorithm and its performance characteristics, we consider two example scenario datasets. The first is a standard phantom measurement which allows us to test our method on a known ground truth. The second dataset is a binaural signal which represents an application of our algorithm workflow to realistic experimental data. 
% Both are critical for determining the overall algorithmic workflow. 
Two evaluation metrics are considered. The first is the sparse sensor signal reconstruction error relative to the original MEG signal. The second is the position/orientation localization error relative to a known or inferred source location, which is a primary goal for general MEG measurements.  
%Both of these are evaluated within the context of the phantom and binaural data.

All data analysis was performed within Matlab and with Brainstorm \cite{tadel11}, which is documented and freely available for download online under the GNU general public license (\url{http://neuroimage.usc.edu/brainstorm}).

\subsection{Sensor selection on phantom signal}
\subsubsection{200 nAm and 200 nAm signals} \label{200_nAm_and_200_nAm_signals}

The first dataset that we use is a measurement performed by John Mosher (Epilepsy Center, Cleveland Clinic Neurological Institute, Cleveland, Ohio, USA) using a 306-channel Elekta Neuromag system (Megin, Helsinki, Finland) with a dry current phantom. The dataset is provided by Ken Taylor and John Mosher on the Brainstorm website (\url{https://neuroimage.usc.edu/brainstorm/Tutorials/PhantomElekta}).

% We note that the 306-channel Elekta Neuromag system consists of 204 gradiometers and 102 magnetometers that measure at different sensitivities. The gradiometers measure at around 100 times their magnetometer counterparts, and thus...

The phantom contains 32 dipole sources located at 4 different depths from 33.9 mm to 63.9 mm, and is a well-established model for validating MEG systems and methods. Three sets of measurements of varying source strengths were provided in Brainstorm: 2000 nAm, 200 nAm and 20 nAm; they correspond to unrealistically high, realistic, and weak source strengths/SNR respectively. We use only the first two sets (2000 nAm and 200 nAm). The dipoles are sequentially activated, and 32 epochs corresponding to each dipole activation are selected from the 2000 nAm and 200 nAm datasets each to test the QR pivoting algorithm for sensor selection. We followed the tutorial provided on the Brainstorm website for signal processing, except the recordings were not resampled to be 100 Hz; they were kept at the original 1000 Hz. Note that from here on, we will refer to these epochs as ``datasets'' still, for convenience. 

\paragraph{Signal reconstruction}

\begin{figure}[ht]
    \centering
    \begin{subfigure}[t]{0.49\textwidth}
        \includegraphics[width=\textwidth]{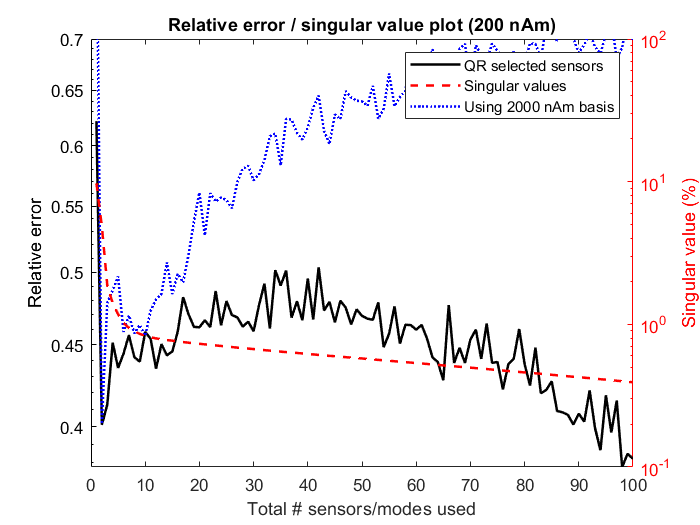}
        \caption{}
    \end{subfigure}
    \begin{subfigure}[t]{0.49\textwidth}
        \includegraphics[width=\textwidth]{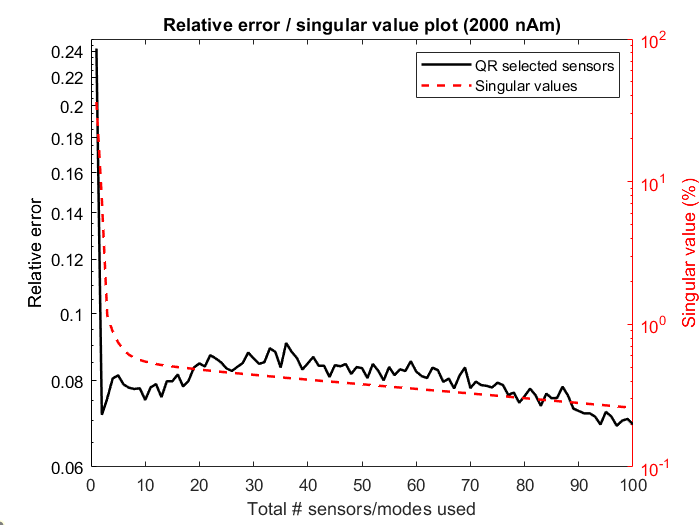}
        \caption{}
    \end{subfigure}
    \caption{(a) Plots for the 200 nAm case; all curves are averaged over all 32 dipolar source results. The red dashed curve shows the mean singular value spectrum and the black curve shows the mean relative error of the reconstructed signals with respect to the original signals. The elbow of the singular value spectrum occurs at approximately 2-10 modes, and we see relative error is minimum in that range as well. The relative error curve increases afterwards, peaking at around 35 modes selected, before decreasing afterwards. The blue dotted curve shows the mean relative error when we use the 2000 nAm SVD basis for signal reconstruction. Again, low relative error is observed between 2-10 modes before it increases, this time peaking much higher at a value above 35 modes. (b) Plots for the 2000 nAm case; the behavior of the curves are similar to the 200 nAm case.}
    \label{fig:phantom_relerr}
\end{figure}

We first performed SVD on each of the 32 datasets for the 2 different source strengths. This is to obtain the left singular matrix that will be used as our basis matrix that we perform pivoted QR on, as well as obtain the singular value spectrum to estimate the underlying number of significant modes. To show that we may estimate the number of modes/sensors to keep from the singular value curve, we kept $r = 1, \dots, n$ modes and did a signal reconstruction for each case with $p=r$ sensors selected by the QR pivoting algorithm, then calculated the relative error of each reconstruction. Let the reconstructed signal with $p=r$ sensors be denoted as $\bm{\Phi}_p'$. The relative error (RE) was calculated via
\begin{align}
    RE_p = \frac{\norm{\bm{\Phi} - \bm{\Phi}_p'}}{\norm{\bm{\Phi}}}
\end{align}
For the 200 nAm and 2000 nAm cases, the mean singular value spectrum and mean relative error over all 32 datasets are shown in Figure~\ref{fig:phantom_relerr}.

Figure~\ref{fig:phantom_relerr} highlights two important aspects of the data: (i) its low-dimensional structure, and (ii) its ability to be reconstructed with limited measurements. Specifically, this figure shows that for both high and low signal strengths, there exists a low-dimensional embedding to global SVD modes. 

From the mean singular value curve, the elbow for both 200 nAm and 2000 nAm cases appear to be at around 2-10 modes. Correspondingly, there is low relative error within that domain, before relative error increases to a peak at around 35 modes. This indicates that it is appropriate to truncate at the number of modes corresponding to the elbow of the singular value curve.

The behavior of the relative error curve is expected and may be explained as follows. The first few significant SVD modes correspond to modes representing meaningful features of the signal (or ``clean'' modes), hence relative error decreases. The subsequent modes correspond to noise/noisy modes \cite{epps19,gavish14}. By including just a few of these noisy modes, the actual random noise in the signal cannot be projected onto them accurately, thus error increases. If we continue increasing the number of noisy modes used however, noise can be more accurately represented gradually, and error starts decreasing again. The observation that the 2000 nAm dataset (with higher SNR) has larger first few singular values also indicates that the first few significant modes are clean and indeed correspond to signals, whereas the subsequent modes correspond to noise.

\begin{figure}[ht]
        \centering
        \includegraphics[width=\textwidth]{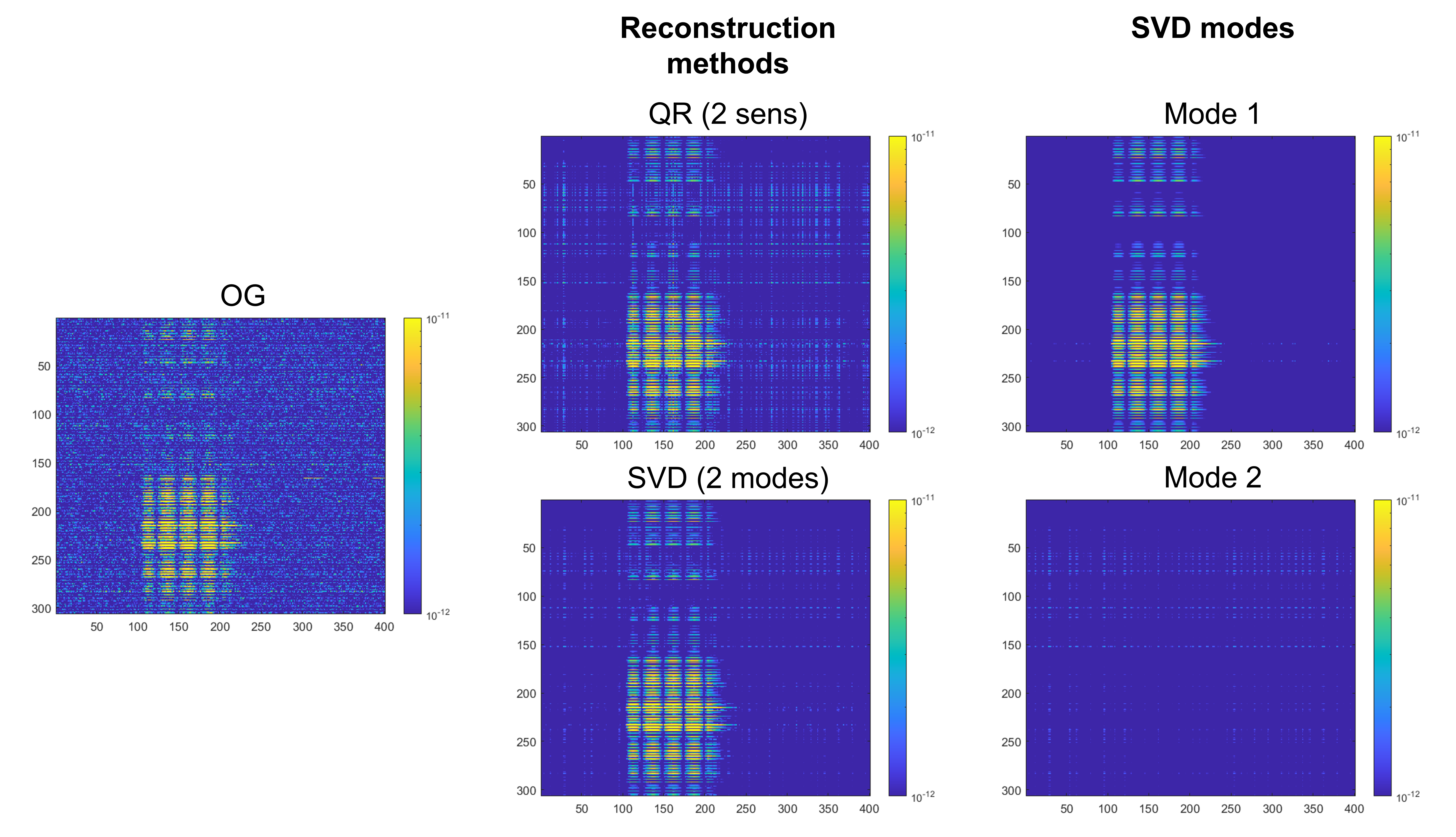}
        \caption{Image plots of the original (OG, left column) and reconstructed (middle column) signals of a (-5.97,0,2.29) cm, 200 nAm phantom dipolar source. The 2-mode 2-sensor QR reconstruction has similar performance to the 2-mode 306-sensor SVD reconstruction, indicating the redundancy of dense sensor arrays. The plots of the first 2 SVD modes on the right column visually shows how the first mode largely resembles the original signal. This illustrates how only the first few significant modes are clean, and exhibits the sparseness of the signal.}
        \label{fig:modesplot}
\end{figure}

To illustrate how very sparse sampling can yield accurate results, we will use the 2-mode QR reconstruction with 2 sensors. Figure~\ref{fig:modesplot} shows the signals of one of the 32 200 nAm dipoles as a color plot. The original signal is on the left (OG), the 2-mode QR and 2-mode SVD reconstructions are in the middle column, and the first 2 SVD modes are in the rightmost column. The 2-mode QR reconstruction using 2 sensors yields comparable results to the 2-mode SVD reconstruction using all 306 sensors, signifying the redundancy of many sensors; we need only 2 optimally-selected sensors that best represent the 2 SVD modes to reconstruct the signal well. We also see that the first SVD mode resembles the original signal strongly, illustrating that it is a clean mode.

\begin{figure}[ht]
    \centering
    \begin{subfigure}[t]{0.4\textwidth}
        \includegraphics[width=\textwidth,keepaspectratio]{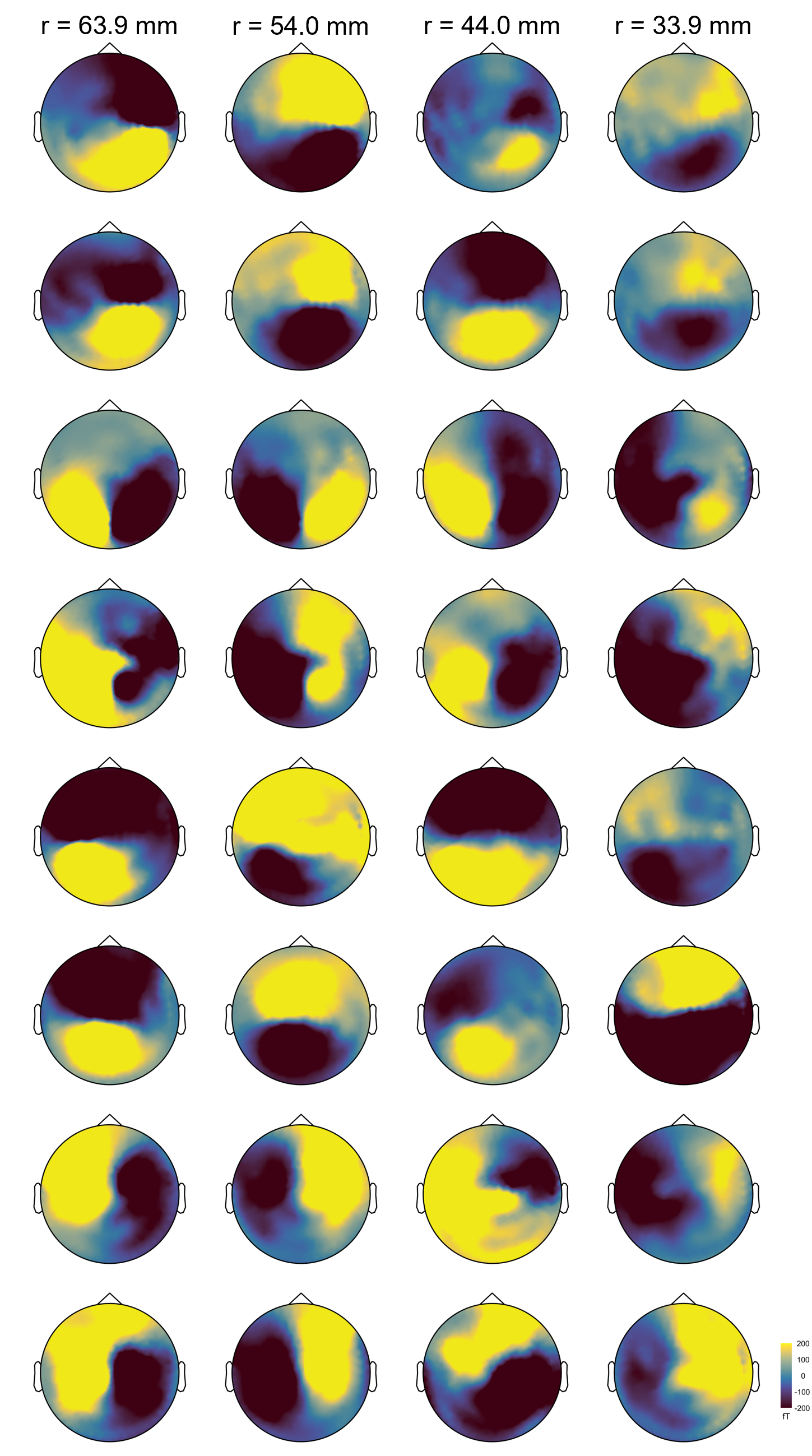}
        \caption{}
    \end{subfigure}
    \qquad
    \begin{subfigure}[t]{0.4\textwidth}
        \includegraphics[width=\textwidth,keepaspectratio]{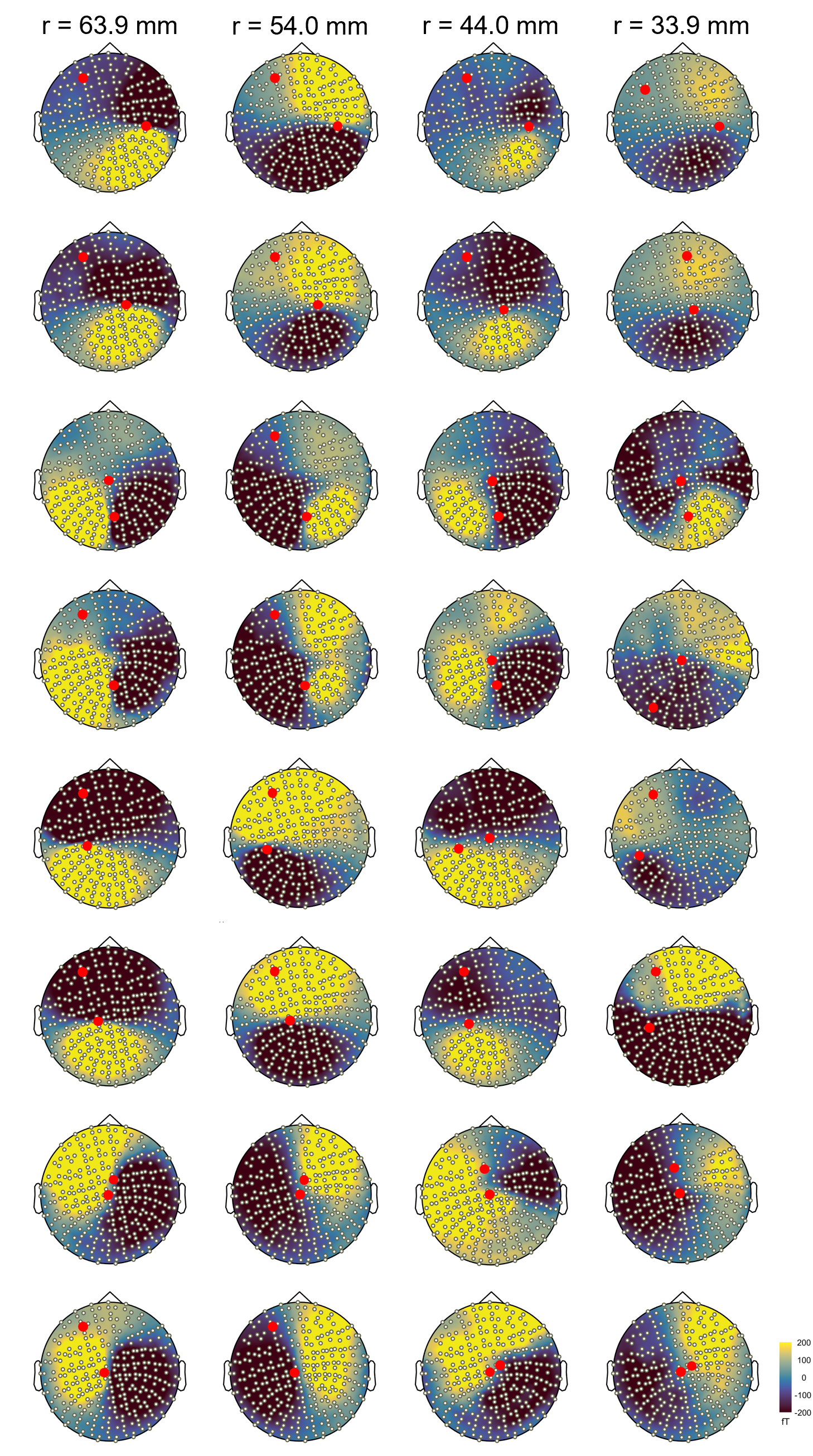}
        \caption{}
        \label{fig:200_fieldsb}
    \end{subfigure}
    \caption{(a) Original signals for each of the 200 nAm 32 phantom dipolar sources. (b) 2-mode QR reconstructed signals, with the 2 selected sensors indicated by the red dots. There is high reconstruction accuracy, and sensor selection appears to be source-dependent.}
    \label{fig:200_fields}
\end{figure}

\begin{figure}[ht]
    \centering
    \begin{subfigure}[t]{0.4\textwidth}
        \includegraphics[width=\textwidth,keepaspectratio]{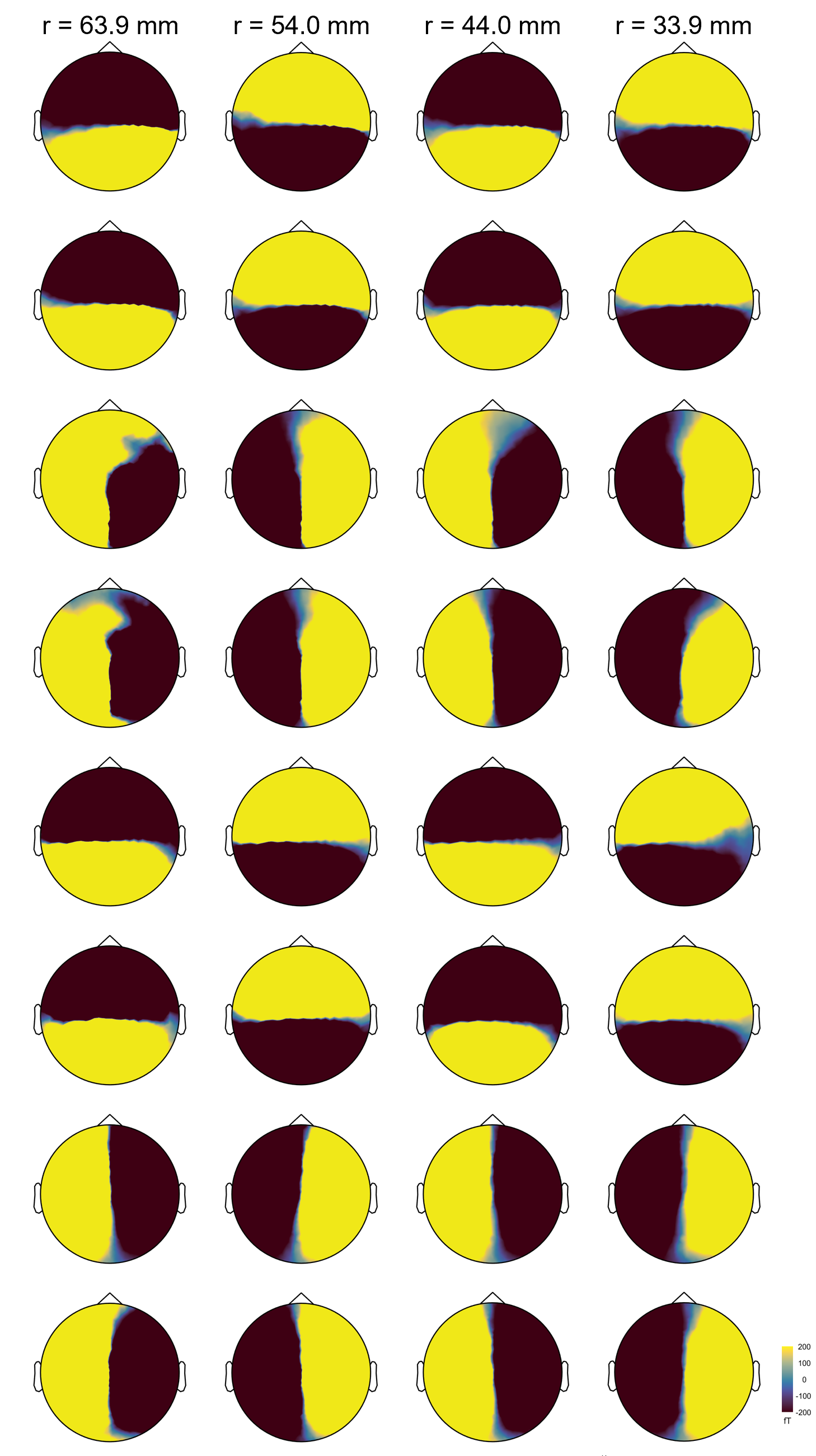}
        \caption{}
    \end{subfigure}
    \qquad
    \begin{subfigure}[t]{0.4\textwidth}
        \includegraphics[width=\textwidth,keepaspectratio]{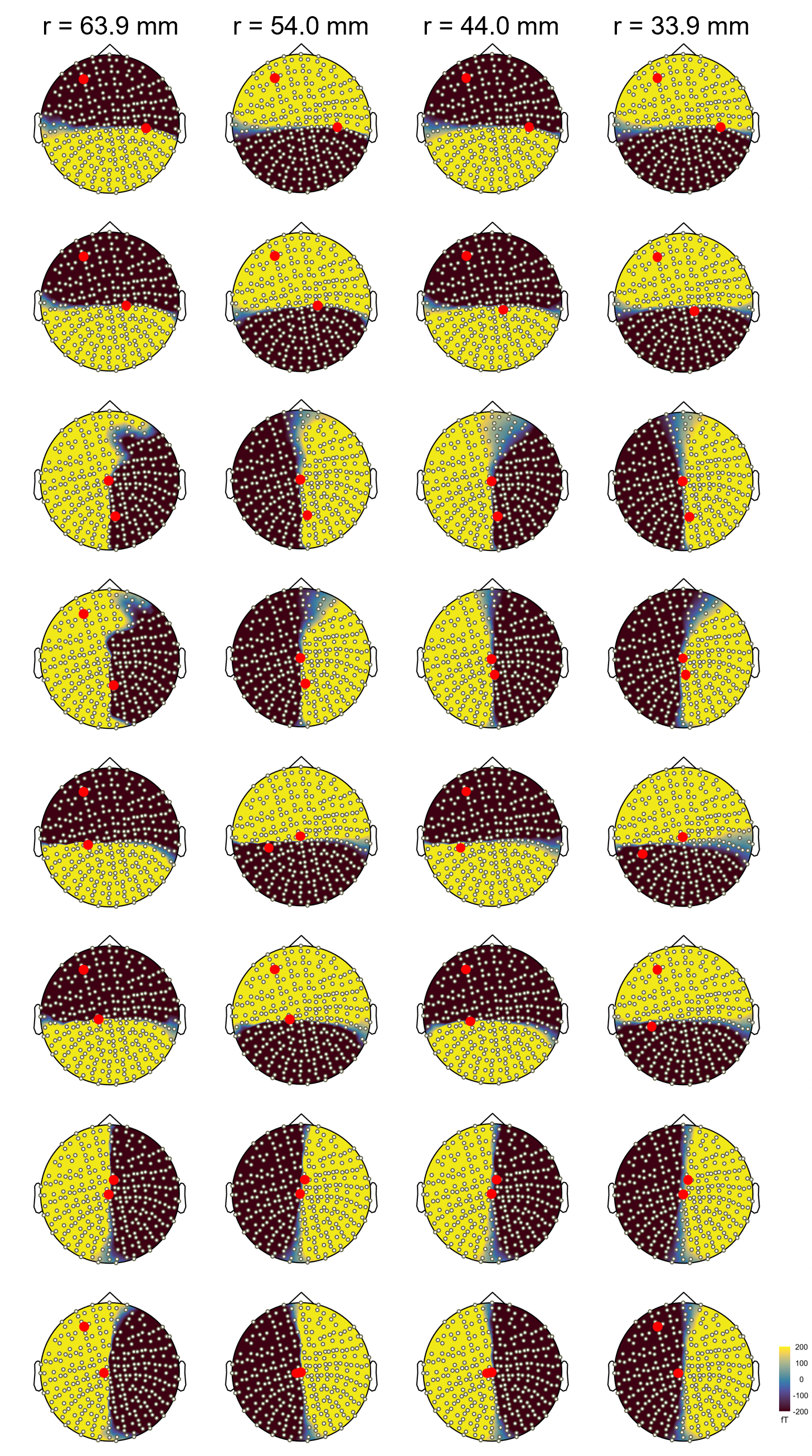}
        \caption{}
        \label{fig:2000_fieldsb}
    \end{subfigure}
    \caption{(a) Original signals for each of the 2000 nAm 32 phantom dipolar sources. Note the more distinct dipolar field patterns due to the higher SNR. (b) 2-mode QR reconstructed signals, with the 2 selected sensors indicated by the red dots. There is high reconstruction accuracy, and sensor selection appears to be source-dependent.}
    \label{fig:2000_fields}
\end{figure}

Figures~\ref{fig:200_fields} and \ref{fig:2000_fields} show the original interpolated scalar magnetic flux field for each of the 32 dipole sources (left), as well as the 2-mode QR reconstructed fields with the 2 selected sensors (right). The 2-mode QR reconstructed fields are very similar to the original, which is in agreement with the low relative error value of Figure~\ref{fig:phantom_relerr} at 2 modes. This indicates that the QR pivoting algorithm works well in selecting optimal sensors for accurate signal reconstruction. The selected sensor positions also indicate that sensor selection is dependent on source location, especially in the tangential direction. At least one sensor close to the vicinity of source activity is always selected, and the 2 sensors selected across the rows of Figures~\ref{fig:200_fieldsb} and \ref{fig:2000_fieldsb} (which have same tangential coordinates but different radial depths) are generally placed in similar locations. As expected, for the stronger 2000 nAm case, the sensor locations are more consistently chosen due to the higher SNR that reveals the true dipolar signal pattern more overtly.

\paragraph{Source localization}

As mentioned, we also consider source localization error. An equivalent current dipole (ECD) fit is provided in Brainstorm, which we utilize at 60 ms for each of the 32 dipole datasets as instructed in the Brainstorm tutorial.

\begin{figure}[ht]
        \centering
        \includegraphics[width=\textwidth]{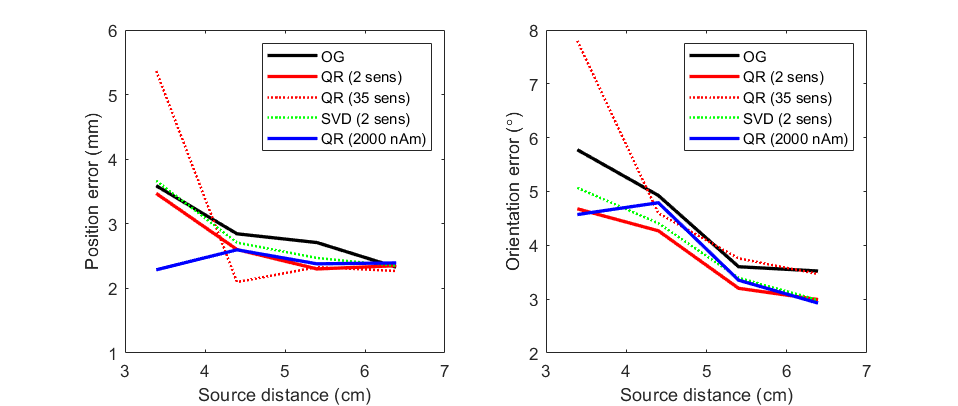}
        \caption{Position and orientation localization errors of ECD fits using 200 nAm signals, namely the original (OG) signal (black curve), 2- and 35-mode QR reconstructed signals (red curves), 2-mode SVD reconstructed signal (green curve), and the QR reconstructed signal using 2000 nAm basis (blue curve). All cases except the case using 2000 nAm basis had higher localization errors for deep sources, especially the 35-mode QR case. This indicates that using more modes when applying the QR algorithm does not guarantee better performance. For more superficial sources, all cases including the 2-mode QR case had similar localization performance, indicating redundancy in sensors.}
        \label{fig:locerr_200_1}
\end{figure}

\begin{figure}[ht]
        \centering
        \includegraphics[width=\textwidth]{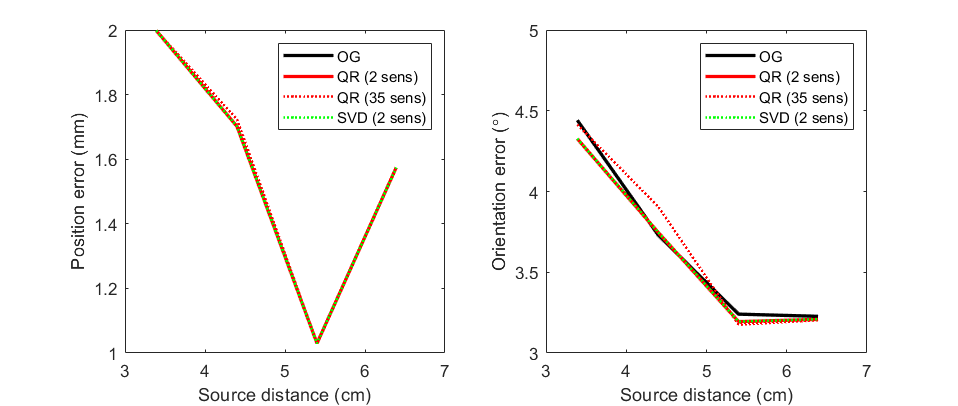}
        \caption{Position and orientation localization errors of ECD fits using 2000 nAm signals, namely the original (OG) signal (black curve), 2- and 35-mode QR reconstructed signals (red curves), and the 2-mode SVD reconstructed signal (green curve). All cases performed identically well due to the high SNR. }
        \label{fig:locerr_2000}
\end{figure}

Figures~\ref{fig:locerr_200_1} and \ref{fig:locerr_2000} show the position and orientation source localization error using QR and SVD methods for both the 200 nAm and 2000 nAm datasets respectively. They are plotted with respect to dipole source depth, so each data point is obtained by averaging the localization error of the $32/4 = 8$ dipoles of same radial distance. The reference is a standard ECD fitting using the original (OG) signal.

\begin{figure}[ht]
        \centering
        \includegraphics[width=\textwidth]{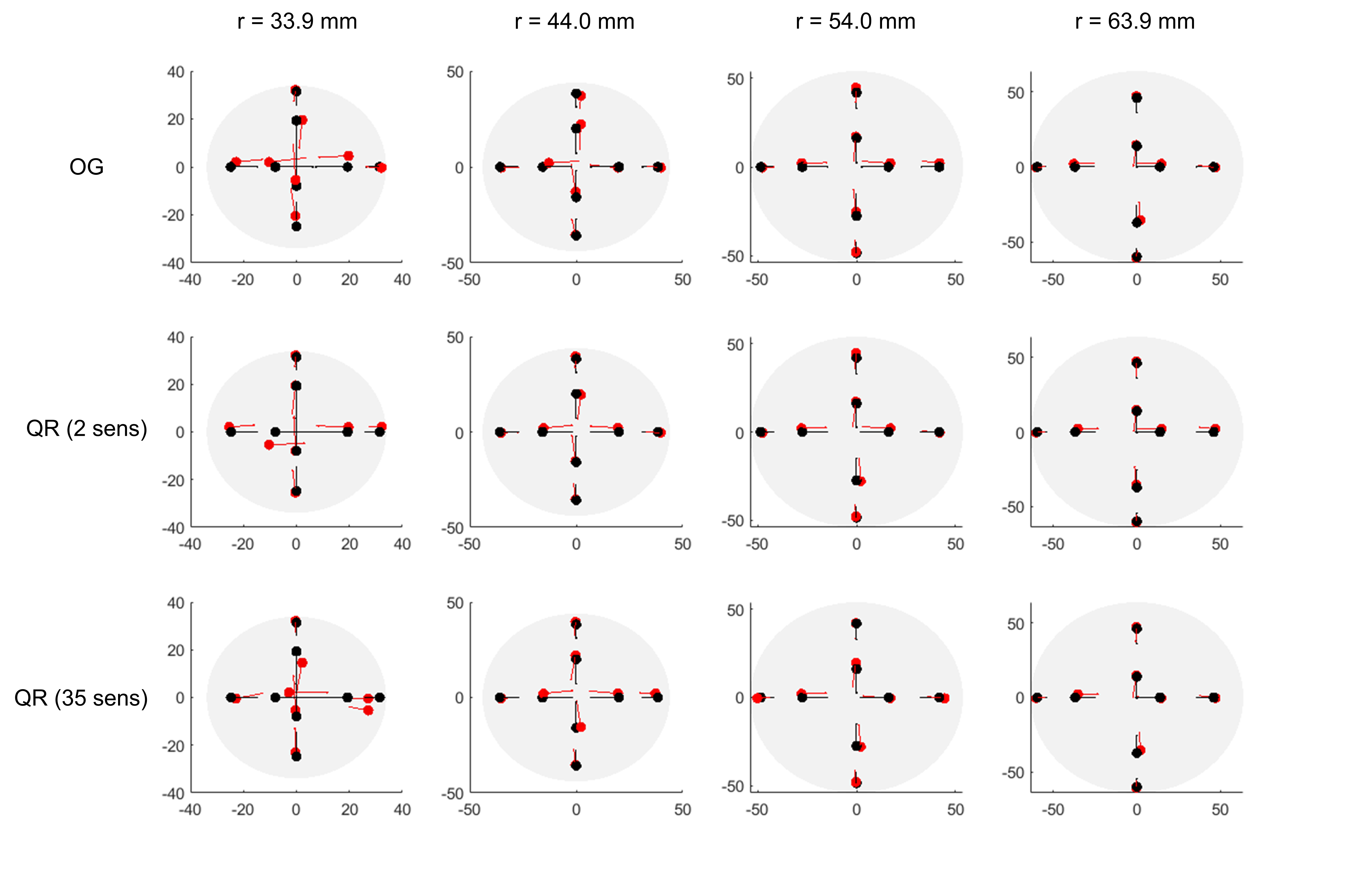}
        \caption{Visualization of ECD fit error when using original (OG) signal, and 2- and 35-mode QR reconstructed signals against the various source depths. Clearly the the localization becomes poorer as sources are deeper, which agree with Figures~\ref{fig:locerr_200_1} and \ref{fig:locerr_2000}.}
        \label{fig:locerr_200_2}
\end{figure}

For the 200 nAm case in Figure~\ref{fig:locerr_200_1}, the 2-mode QR localization is comparable to both the 2-mode SVD and OG localizations, again indicating the redundancy of many sensors if working in the SVD basis. The 35-mode QR localization however suffers from poorer localization at deep source distances, which is to be expected given the higher signal reconstruction relative error as seen in Figure~\ref{fig:phantom_relerr}. We also notice that deeper sources have poorer source localization. This can be understood in the spherical harmonics decomposition framework, as briefly mentioned in Section~\ref{introduction} \cite{taulu05}. High spatial frequency components of the signal that allows for higher resolution decay quicker than low spatial frequency components. More precise source localization can be achieved with higher signal resolution. Since deeper sources are further from the sensors, the signals recorded contain the lower-frequency components only, resulting in poorer localizations than those of more superficial sources. For easier visualization, a plot of the physical locations and orientations of the OG and QR reconstructed dipoles is shown in Figure~\ref{fig:locerr_200_2}. Each column corresponds to the different source depths, and we see that the deepest sources at $r=33.9$ mm indeed do suffer from poorest localization, consistent with Figure~\ref{fig:locerr_200_1}. Note that the plots are viewed top-down from the $z$-axis.

For the 2000 nAm case in Figure~\ref{fig:locerr_2000}, the unrealistically high SNR provided near-identical good performance as expected, with $<2$ mm and $<4.5^\circ$ position and orientation errors relatively. Again, deep sources suffer from poorest localization.
% although we note that this may be a false positive result. The position error values may be simply fluctuating about its minimum attainable range, i.e. the ECD fit may be yielding fits at its highest possible resolution for all source distances.

% Of course, one cannot set up truly dipolar current sources in real life due to a violation of the continuity equation. However, it has been shown in \cite{ilmoniemi09} that for a spherical conductor model, we may set up a current triangle loop that produces a magnetic field equivalent to a dipolar source. Since radial edges with respect to some chosen origin give zero magnetic field contribution, we may choose the origin of the system to coincide with the center of the sphere and accurately model a purely dipolar contribution with the last non-radial edge. This principle is used to set up the phantom. 

\subsubsection{2000 nAm basis on 200 nAm}

In reality, if we want to have a system of $p$ sensors, we will need to obtain information about where to place them from either a simulation or prior datasets measured with all sensors. The basis matrix from the full-state signal is also required to obtain a reconstruction from the $p$ measurements. As mentioned in the previous section, since sensor selection is dependent on source location, the sources for these simulations/prior recordings will necessarily need to be approximately the same as what we expect to have for the $p$ sensor set-up or robust enough so that the sensor placements are indeed nearly optimal. 

Here, we verify that if we apply SVD and QR pivoting on a simulation/prior dataset of all sensors, then use the sensor locations selected and basis obtained to reconstruct the full-state signal from a $p$-sensor measurement with similar source configuration, that accurate source reconstruction is still attainable. The simulation/prior dataset that we use is the 2000 nAm dataset with all 306 sensors, which mimics an overly-ideal simulation with high SNR. The $p$ sensor locations selected from the QR pivoting of the 2000 nAm dataset is used to define the $p$-sensor 200 nAm measurement that we want to perform a full-state signal reconstruction and source localization on.

The results are shown in the blue curves of Figures~\ref{fig:phantom_relerr} and \ref{fig:locerr_200_1}. In Figure~\ref{fig:phantom_relerr}, we see that the relative error for signal reconstruction is comparable to using the 200 nAm basis (black curve) up to about 10 modes as expected, since these first few significant modes represent the identical clean signal. However, the relative error increases rapidly afterwards. This may be explained by how the subsequent modes for the 2000 nAm basis are cleaner and do not represent the noise of the 200 nAm signal as well. If more 2000 nAm modes are included, the error eventually decreases since the noise is gradually represented better as explained in Section~\ref{200_nAm_and_200_nAm_signals}, but now significantly more number of modes are required than if we used the 200 nAm basis itself. Note that if the plot is extended to beyond 100 sensors/modes, the blue curve eventually decreases. The 2-mode 2000 nAm basis reconstruction of the 200 nAm data performs comparably to the original source localization methods, as seen in Figure~\ref{fig:locerr_200_1}. It is interesting to note that deep sources do not suffer from higher position localization error in this case, which indicates that using a high SNR simulation basis may have noise suppression capabilities when reconstructing noisy data. Intuitively, this may be understood by the fact that we are interpolating data in a sense from the $p$ sparse sensor measurements based on the clean basis, thus no noise is being reconstructed.

\subsection{Sensor selection on binaural stimulation signal}

To show the validity of QR pivoting algorithm for sensor selection on an realistic brain measurements, we consider now consider a dataset of binaural stimulation. It is a measurement by Elizabeth Bock, Peter Donhauser, Francois Tadel and Sylvain Baillet (MEG Unit Lab, McConnell Brain Imaging Center, Montreal Neurological Institute, McGill University, Canada) using a CTF 275 system with 274 axial gradiometers (\url{https://neuroimage.usc.edu/brainstorm/DatasetIntroduction}).

Again, we followed the Brainstorm tutorial to identify the relevant epoch/dataset with dipolar-like activity for analysis. Like before, SVD was performed on the dataset, followed by QR pivoting on the SVD basis for sensor selection. Signal reconstruction and ECD source localization were then done at 90 ms of the epoch.

\begin{figure}[ht]
    \centering
    \begin{subfigure}[t,keepaspectratio]{0.49\textwidth}
        \includegraphics[width=\textwidth]{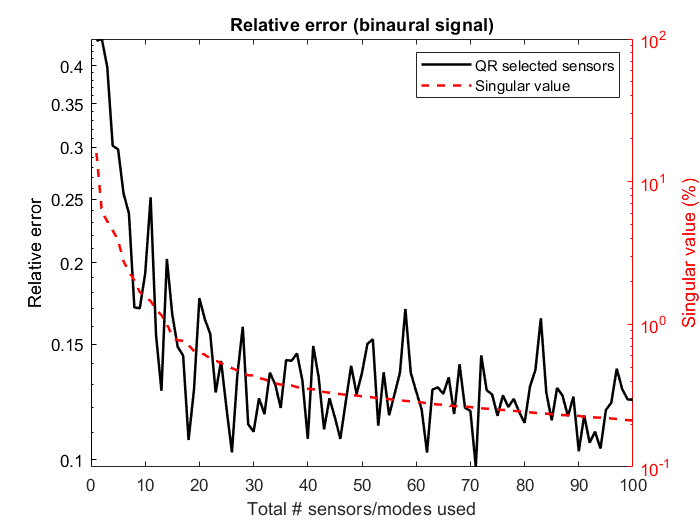}
        \caption{}
        \label{fig:binaural_err_a}
    \end{subfigure}
    \qquad
    \begin{subfigure}[t,keepaspectratio]{0.4\textwidth}
        \includegraphics[width=\textwidth]{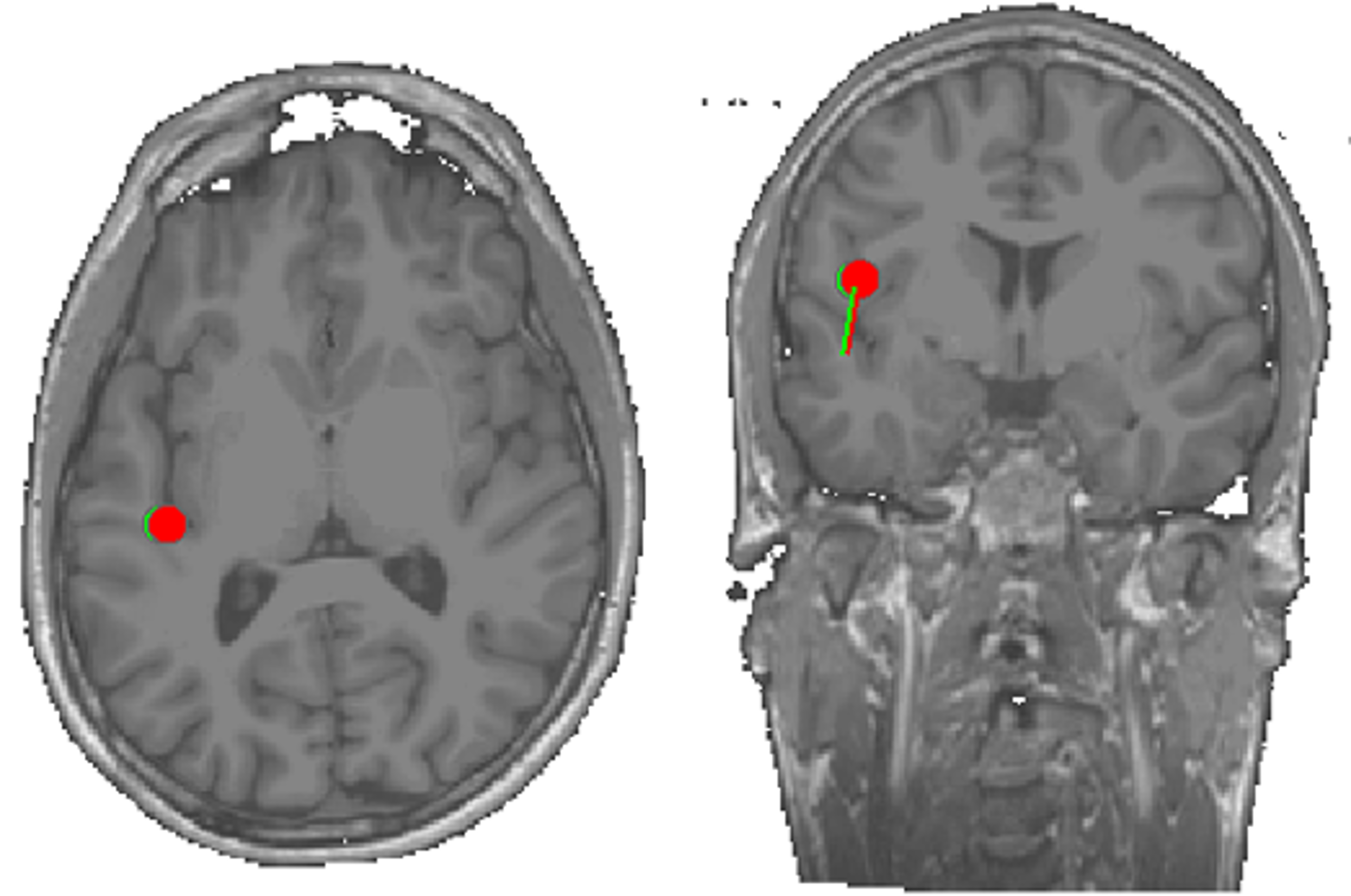}
        \caption{}
        \label{fig:binaural_err_b}
    \end{subfigure}
    \caption{(a) Relative error plot for binaural signal relative to the original signal (black curve). The singular value plot is also shown (red curve). The relative error plot follows the singular value spectrum as expected, indicating number of modes/sensors can be selected from the singular value spectrum of a dataset. (b) Superior (left) and posterior (right) view of the head MRI that shows the dipole localizations using 274-sensor original signal (red) and 30-mode QR reconstructed signal (green). The performances are near-identical, indicating sensor redundancy.}
    \label{fig:binaural_err}
\end{figure}

The results are shown in Figure~\ref{fig:binaural_err}. In Figure~\ref{fig:binaural_err_a}, the singular value spectrum now does not have a distinct elbow and decays gradually, which makes sense given the more complex source configuration and higher noise levels. Due to the noisy signal, no increase then decrease of relative error is observed when we gradually include more modes as well. We estimate that 30 sensors/modes are sufficient for accurate reconstruction. The relative error curve when using $p$-mode QR reconstruction shows that it follows the singular value curve, indicating again that selecting the number of sensors/modes based of the singular value curve is valid. An ECD fit was then done using both the original data and the 30-mode QR reconstructed signal; the result is visually presented in Figure~\ref{fig:binaural_err_b}. The fit using original data is displayed as a red dipole, whereas the fit using 30-mode QR reconstructed signal is displayed as a green dipole. Their localization differences are small (hence the large overlap between the two dipoles), with 1.4142 mm position difference and $0.7966^\circ$ orientation difference. This indicates that the localization performance using just 30 sensors is comparable to using 306 sensors.
 
\section{Conclusion}

In this paper, we demonstrated that for MEG datasets with spatially localized sources, we may exploit its low-rank SVD representation to select a small number of sensors that will still be able to reconstruct the full-state signal well, as well as localize sources accurately. The elbow of the singular value spectrum from the SVD may be used to determine the number of modes/sensors to truncate at, then the QR pivoting algorithm may be performed on the SVD left-singular basis to greedily determine the near-optimal sensor positions.

From our results, we see that sensor selection is source-dependent. Therefore, the prior simulations or datasets that we perform SVD and QR pivoting on in order to determine optimal sensor location are required to have source configurations similar to what we realistically expect to measure. For more robust source localization, it may be possible to train a neural network to perform sensor selection with a training set that contains simulations varied source configurations.

In addition to selecting a few optimal sensors from current MEG sensor array layouts, the SVD and QR pivoting pipeline has other advantages, including sensor array design. If we utilize a simulation that has a large number of sensors, sensor selection can determine where to place sensors with high resolution (given a specific source configuration). Moreover, full-state reconstruction with limited sensors has additional capabilities of interpolating/extrapolating to regions inaccessible for measurements, for example areas obstructed by electronic components. The use of less sensors may be more computationally efficient \cite{nenonen22}, although the possible extra step of having to run prior simulations is a factor to consider.

One possible limitation sensor sparsity is the increased importance of coregistration errors \cite{zetter18}. Without a full-head coverage, misplacing the optimally-selected sensors may render them non-optimal and result in inaccurate full-state reconstruction. We also acknowledge advantages to having sensor redundancy; having a dense sampling of the head gives a more comprehensive measurement of head activity without the need for signal reconstruction. The signal reconstruction of from sparse data relies heavily on the basis choice which is obtained from simulation; if the simulation or prior dataset does not resemble the actual source configuration of interest, large reconstruction errors may occur. Practically, more sensors also mean more backups in case of sensor failure.

As a final note, the sensor selection pipeline presented in this paper may also be extended to EEG, which has a similar mathematical formulation to MEG.

% more general training set means more robust applications to more general sources in diff locations
% shannon info limit due to snr (nenonen07)
% scan head for accurate sensor locations for accurate simnulations also (e.g. returning patients?)

\section*{Acknowledgements}

JNK acknowledges support from the Air Force Office of Scientific Research (FA9550-19-1-0386 and FA9550-19-1-0011). ST is supported by the Bezos Family Foundation and the R. B. and Ruth H. Dunn Charitable Foundation.

% \appendix

\bibliographystyle{IEEEtran}
\bibliography{references.bib}

\end{document}